# Metasurface-Enhanced Transparency

Christopher M. Roberts, Viktor A. Podolskiy

*Department of Physics and Applied Physics, University of Massachusetts Lowell, One University Avenue, Lowell, Massachusetts 01854, USA*

We consider the problem of light transmission from a high refractive index medium into a low index environment. While total internal reflection severely limits such transmission in systems with smooth interfaces, diffractive metasurfaces may help out-couple light that enters an interface at blazing angles. We demonstrate that the profile of the structured interface can be numerically optimized to target a particular emission pattern. Our study suggests that while metasurfaces can help to out-couple light from a range of incident directions, there exists a universal limit for total transmission efficiency that depends only on the dielectric properties of the two materials and is independent of the profile and the composition of the metasurface coupler.

Efficient light coupling is important for the design and development of detectors, emitters, and routers of light. For years, smooth anti-reflection (AR) coatings have been used to decrease reflection and thus increase transmission via destructive interference or a refractive index gradient [1]. More recently, biomimicry-based Moth's eye films have emerged as a new class of anti-reflection elements that rely on arrays of tapered nanostructures to provide an index gradient and to reduce reflection [2,3]. However, the majority of existing AR-coatings aims to minimize the reflection of light that couples from (low-index) air to a relatively high-index medium (glass, semiconductors, etc.). As result, advances in optically-smooth AR coatings cannot be applied to the reciprocal geometry, where the light couples from a high-index media [such as (O)LEDs or scintillators] into lower-index surroundings, which is severely limited by total internal reflection. In this work, we consider diffractive AR coatings that can overcome the limitations of total internal refraction [4]. We show that it is possible to optimize transmission through an interface for a specific emission pattern and present several designs of optimal AR coatings based on diffractive metasurfaces[7–13]. More importantly, our extensive numerical study and analytical arguments suggest the existence of a fundamental limit to the overall interface transparency, which depends only on the dielectric properties of the two surrounding materials and does not depend on either the composition or the geometry of the interface itself.

The transmission of light through a smooth interface is governed by the well-known Fresnel equations [4] that combine the light's polarization, incident angle, and the dielectric properties of the two materials. As illustration, Fig.1 shows transmission through an interface between air ($n = 1$) and material with index $n = 3.3$ (that represents index of a typical semiconductor [5]). It is seen that when light couples from the air side, a substantial portion of the light is reflected due to an impedance mismatch (Fig.1a). Similar parasitic reflection exists for the inverse configuration where the light exits a high-index medium in almost-normal direction (Fig.1b). Conventional AR coating designs aim to minimize this impedance mismatch and thus maximize light coupling for almost-normal incidence.

When light is incident on the interface from a high-index material, transmission is limited not only due to an impedance mismatch but also due to total internal reflection (higher angles in Fig.1b). This phenomenon cannot be avoided with any translationally-invariant AR coating design [4], and thus it affects performance of any application where non-paraxial light is generated inside a high-index medium.

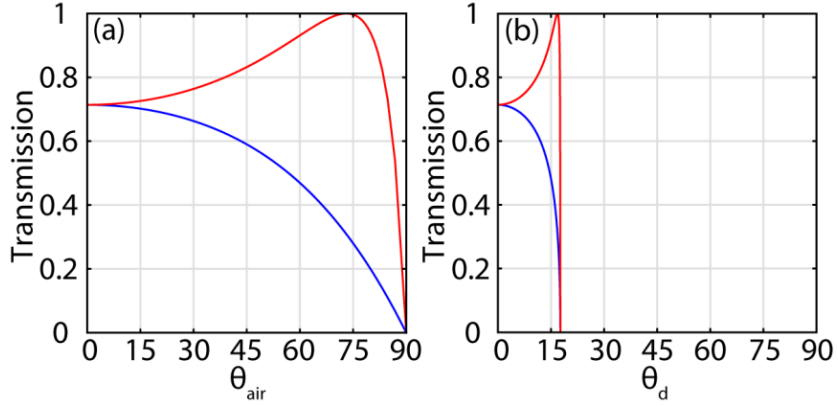

*Figure 1: Total transmitted power between a dielectric ($n = 3.3$) and air ($n = 1$). TM- and TE- polarized light is represented by red and blue lines respectively; panel a) represents transmission from air to dielectric, panel b) illustrates transmission from dielectric into air*

To quantify transmission of light through the interface we represent the incident light as a linear combination of plane waves, parameterized by the in-plane component of their wavevector, $k_x$. The total transmission through the interface, playing the role of a Figure of Merit ($FOM$) for this work is defined as

$$FOM = \sum_{i=-\infty}^{\infty} \int_{-\frac{\omega n_1}{c}}^{\frac{\omega n_1}{c}} T_i(k_x) dk_x \qquad (1)$$

where $T_i(k_x)$ represents the percentage of energy flux that enters the interface with in-plane wavevector $k_x$ and is diffracted to the diffraction order $i$ the across the interface. For the material combination discussed above, only ~24% of light generated inside the high-index medium makes it to low-index surrounding through a smooth interface, while the remaining power is reflected back into the high-index material. Conventional anti-reflection coatings can minimize reflection (and thus maximize transmission) for small incident angles, below the critical angle of total internal reflection $\theta_c = \sin^{-1} n_2/n_1 \simeq 17°$, potentially raising $FOM$ to $FOM_{max} = n_2/n_1 \simeq 0.3$. Note that despite the enhancement in total transmission, light that enters the interface at $\theta > \theta_c$ never makes it outside the dielectric. Therefore, the transmission enhancement based on conventional AR technology cannot, even in principle, address the needs of applications that require the detection of light generated at all angles to the interface, for example, Cherenkov-based scintillators [6].

In contrast to smooth interfaces, diffractive surfaces do not conserve the in-plane component of the wavevector, and can thus be used to out-couple light that enters the interface at sufficiently high incident angles. In this work, we focus on binary gratings that can be fabricated with conventional lithographic techniques which belong to an emerging broad class of structured materials with engineered diffraction, known as metasurfaces [7–13]. Previous research on metasurfaces has focused on their applications to control the direction of the light flow. While some researchers have analyzed utility of metasurfaces to enhance transmission of light *into* a high-index medium[14,15], to our knowledge virtually no research has been done to understand the benefits of diffractive optics to enhance the *efficiency* of transmission of light out of a high-index medium.

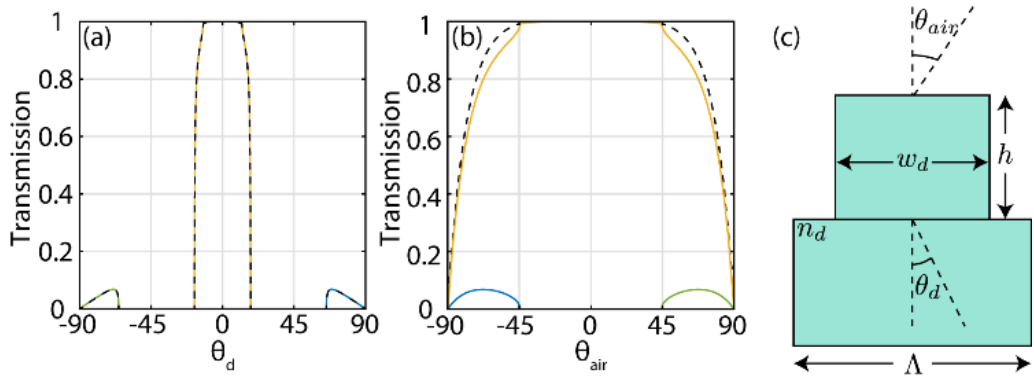

*Figure 2: Optimal extraction of light from high-index dielectric for the grating optimized with a particle swarm technique ($\Lambda = \lambda_0/4$, $f = 0.65$ and $h = 0.13\lambda_0$). Yellow, green, and blue lines represent specular TM transmission as well as diffraction into $m = +1$, and $m = -1$ orders, respectively. The total transmission is shown as a dotted black line. Panels (a) and (b) represent the same process as seen from the dielectric (a) and from air (b). Panel (c) illustrates the geometry of a single unit cell for the optimization study.*

To comprehensively assess the perspective of using diffractive interfaces to solve the problem of light coupling, we have performed three different optimization studies. Overall, we have assumed that the diffractive interface has some (possibly large) period $\Lambda$, and height $h$. In all calculations, our in-house implementation of Rigorous Coupled Wave analysis (RCWA)[16,17] was used to optimize internal structure of the grating, as well as parameters $\Lambda$ and $h$ to maximize $FOM$ in the presence of several constraints.

In the first, simplest study, we considered a simple periodic grating that was parameterized by $\Lambda, h$ and fill fraction $f$ ($w_d/\Lambda$), and used particle swarm optimization technique to optimize the values of these three parameters for a fixed vacuum wavelength $\lambda_0$. Particle swarm optimization [18,19] is a bio-inspired optimization process which uses swarm dynamics to efficiently sample a large global parameter space.  In the beginning of the process, a plethora of parameter tuples (known as "particles") are placed around the global parameter space; coordinates in the space serve as values of parameters being optimized.  At each successive step, each "particle" is endowed with a velocity which defines the next point in the parameter space to be checked.  The velocity of each "particle" is influenced by its local knowledge of the best solution, as well as the global best solution known by all the "particles" in the swarm.  This search technique exhibits a form of swarm intelligence similar to the swarm dynamics of a flock of birds in flight or a school of fish in the sea. [18,19]

For this particular optimization problem, the parameter space was restricted to $0.1\lambda_0 < \Lambda < 10\lambda_0$ and $0 < h < \lambda_0$, to ensure that the resulting metasurface can be fabricated with current lithographic techniques. The optimized $FOM \sim 29.2\%$ was found for a grating with parameters $\Lambda = \lambda_0/4$, $f = 0.65$, and $h = 0.13\lambda_0$. Transmission of this structure is illustrated in Fig.2. When compared to the transmission of a smooth AR-free interface, it is seen that the source of transmission enhancement is two-fold. First, due to the subwavelength $w_d$, the metasurface operates like a moth-eye system, essentially creating a gradient of effective index at the interface, and maximizing the transmission within the critical cone to nearly unity as seen in Figure 2 . In addition, the grating opens two extra, 1st order diffracted channels that allow out-coupling the waves that enter the interface at the glancing angles. Interestingly, the total transmission (that represents the sum of the specular and diffractive waves out-coupled to the same angle in vacuum) never exceeds unity, which indicates that diffraction does not "add" to transmission enhancement of moth-eye process.

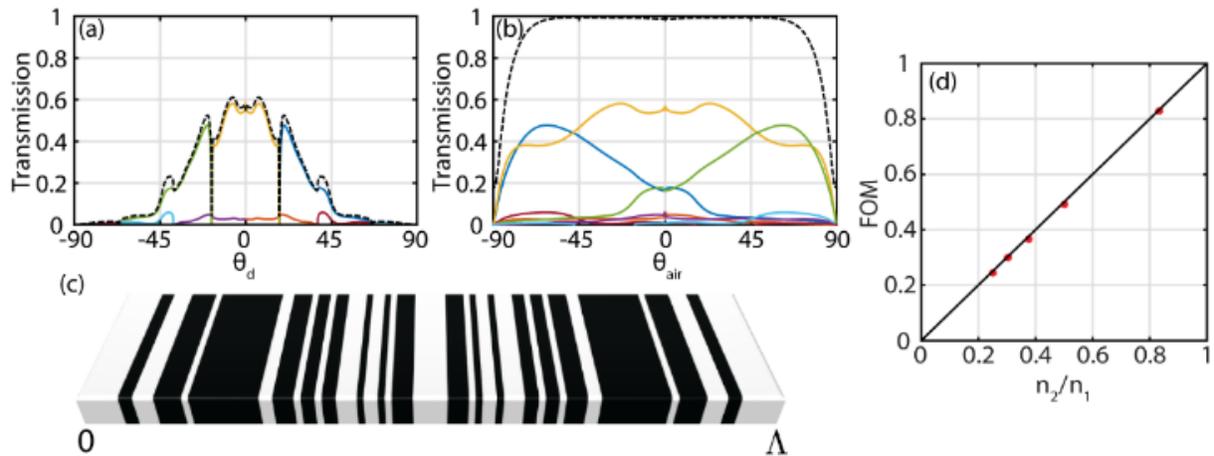

*Figure 3: (a,b) Optimized transmission from the high-index media into low-index environment, as seen from dielectric (a) and air (b), obtained with genetic algorithm for $\Lambda = 0.99\lambda_0$, $h = 0.44\lambda_0$, and $N = 101$. Lines of different color represent contributions from individual diffraction orders (yellow, green, and blue lines correspond to specular, $m = +1$, and $m = -1$ channels); dashed black line represents total transmission. Panel (c) illustrates profile of the optimal metasurface, with white and black regions corresponding to air and high-index dielectric respectively. (d) Maximum total transmission through the interface [see Eq.(1)] for different substrate/superstrate index ratios; line and dots represent Eq.(2) and results of optimization studies, respectively.*

In the next optimization study, we considered more complex binary diffractive structures, represented as a set of $N$ pixels, $\Lambda/N$-wide each. Thus, the grating profile was parameterized by a binary string with 0[1] representing air [high-index dielectric] respectively. Fixing the total pixel number $N = 101$, the profile of the grating (along with overall grating width and height) was optimized with a genetic algorithm, similar to what has been described in our previous work [20]. 'Non-binary' parameters were encoded into the genetic algorithm by prepending their IEEE-754 representations to the gratings binary string representation. Additionally, a reflection symmetry on the grating was imposed, to ensure a symmetric angular transmission distribution. Our studies show that the added symmetry constraint does not fundamentally affect the resultant figure of merit. The study yielded multiple diffractive structures, all having similar performance.

A typical result of such genetic optimization is shown in Fig.3. Notably, the efficiency of the resultant grating (ranging between 29.2 … 29.6%) is only marginally better than the efficiency of the simple periodic structure described above. At the same time, it is seen that more complex gratings provide a more uniform out-coupling profile that is not dominated by small incident angles. As before, despite multiple diffractive orders contributing to the transmission to the same out-coupling angle, the latter parameter never exceeds unity.

The above studies were repeated for different combinations of the two materials. Binary metasurfaces composed from the high-index dielectrics, as well as from lossy dielectrics and plasmonic media, were analyzed. In all these studies, the transmission into a target angle never exceeded unity, regardless of the relationship between the composition of the metasurface and material properties of the surrounding materials. Lossy interfaces yielded lower transmission than their lossless counterparts. The existence of the upper limit of transmission into a given angle from multiple diffraction channels represents the main result of this work.

The upper limit on the angle-dependent transmission implies the limit on the total transmission of light through a structured interface. For example, the maximum FOM introduced above can be calculated as a ratio of the Poynting-flux integrals (parameterized by the in-plane wavenumber). For transparent isotropic media we obtain

$$FOM_{max} = \frac{n_2}{n_1} \qquad (2)$$

To verify the validity of Eq.(2), we repeated the optimization studies for various combinations of refractive indices of substrate and superstrate, as well as for different materials comprising the metasurface layer. As seen in Fig.3(d), results of these extensive optimizations are in excellent agreement with Eq.(2).

Having a strict limit on the overall transmission allows optimization of the metasurface that aims to equalize transmission from within a high-index medium into low-index surroundings. An example of such structure is shown in Fig.4.

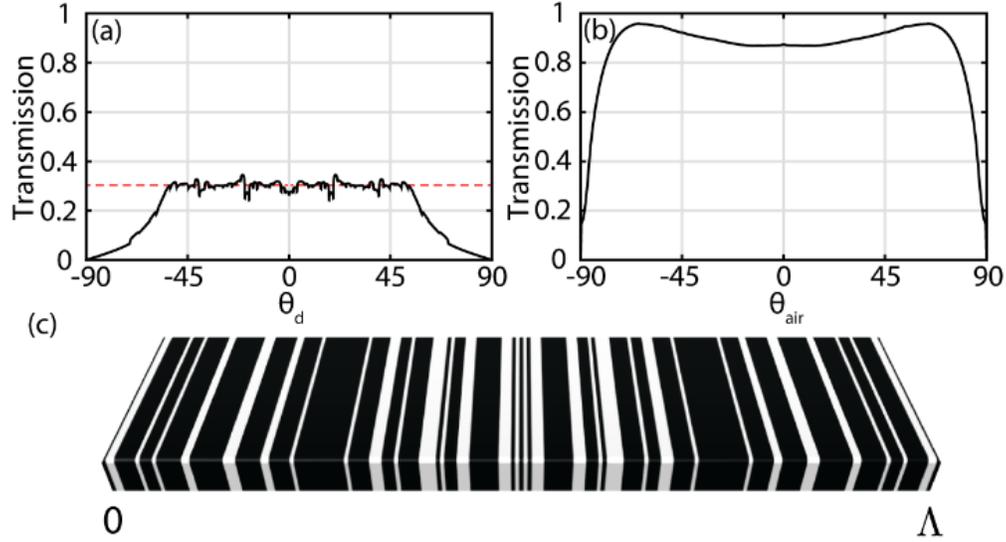

*Figure 4: Nearly angle-independent transmission from high-index into low-index media with diffractive metasurface. (a,b) Total transmission for TM –polarized light as seen from dielectric (a) and air (b) side; black line represents total transmission; red line represents optimization target; $\Lambda = 9.93\lambda_0$, $h = 0.31\lambda_0$, $N = 201$; the minimum feature size $\simeq 0.049\lambda_0$; panel (c) shows the profile of optimal metasurface; black and white regions represent dielectric and air, respectively*

To summarize, we explored the perspectives of diffractive interfaces (metasurfaces) for optimizing the coupling of light between materials with different refractive indices. Several designs of metasurfaces which targeted different emission patterns within high-index medium were presented. Overall, it was found that, independent of interface's composition, the total transmission into any single out-coupled angle cannot exceed unity (transmission is inherently lower for lossy interfaces), which yields a fundamental limit on the efficiency of coupling of mono-chromatic light between two different materials. Although we have considered a (quasi-)planar interfaces between two isotropic media, the results of this work can be readily generalized to applications requiring where the out-coupled light is to be collimated within acceptance angle $\alpha$, where $FOM_{max} = (n_2/n_1)\sin\alpha$, as well as to anisotropic media[4]. The results can serve as starting point to understand limits of transmission in non-planar geometries where the overall transmission is expected to become a function of position of the source within a high-index medium.

This research was partially supported by the NSF-MWN program (grant #DMR-1209761) and NSF-DMREF program (grant #DMR- 1629330)